\documentstyle[prl,aps,tighten,multicol,epsfig]{revtex}

\begin{document}
\draft

\psfull

\title{Long-range effects in granular avalanching} 

\author{Pablo M. Gleiser, Sergio A. Cannas\cite{auth2},
Francisco A. Tamarit\cite{auth2}}

\address{Facultad de Matem\'atica, Astronom\'\i a y F\'\i sica, 
         Universidad Nacional de C\'ordoba, Ciudad Universitaria, 5000 
         C\'ordoba, Argentina}

\author{B. Zheng}

\address{Universit\"at-Halle, 06099 Halle, Germany}

\date{\today}
\maketitle

\begin{abstract}
 We introduce a model for granular flow in a one-dimensional rice pile that incorporates
 rolling effects through a long-range rolling probability for the individual
rice grains proportional to $r^{-\rho}$, $r$ being the distance traveled by a
grain in a single topling event. The exponent $\rho$ controls the average
rolling distance. We have shown that the crossover from power law to stretched
exponential behaviors observed experimentally in the granular dynamics of rice
piles can be well described as a long-range effect resulting from a change in
the transport properties of individual grains. We showed that  stretched
exponential avalanche distributions can be associated with a long-range regime
for $1<\rho<2$ where the average rolling distance grows as a power law with
the system size, while power law distributions are associated with a short
range regime for $\rho>2$, where the average rolling distance is independent
of the system size.
 
 \end{abstract}


There is nowadays a massive evidence of scale invariant phenomena in nature. They appear in 
systems as diverse as geological ({\it e.g.,} earthquakes), climatic ({\it
e.g.,} atmospheric turbulence), granular flow ({\it e.g.,}  rice piles, the
topic of this work) and biological ({\it e.g.}, biological evolution, cell
growth, among many others). In many of these cases no particular tunning is
perceived. In 1987 Per Bak and colaborators \cite{Bak} advanced the hypothesis
 that this is so because the microscopic dynamics of the system makes it to
spontaneously evolve towards a critical, scale-invariant state. This is known
today as {\it self-organized criticality} (SOC). To illustrate the idea they
introduced a 
  sand pile model \cite{Bak} that quickly became the paradigm ofSOC models.
   In this model a pile is gradually built by adding individual
grains one by one into an open system. After a transient period the pile
reaches a stationary critical state, where the slope of the pile fluctuates
around a stationary value. At this stage avalanches of all possible sizes are
observed, giving a power law avalanche size distribution.
This is the SOC state. The experimental search for power laws in the avalanche
size distribution in real granular systems presented a great challenge
\cite{Held,Jaeger,Evesque,Nagel,Bretz}.  Some  years ago the Oslo group 
\cite{Frette2,Christensen} found evidence of SOC in controlled experiments on
the granular dynamics of rice piles.  In 1996 Frette {\it et al} \cite{Frette}
performed an experiment where elongated rice grains were added slowly in a
narrow gap between two plates. They found that the avalanche size distribution
for grains with large aspect ratio presents a power law behavior, while a
stretched exponential behavior is observed for rounder grains. These
experimental results showed that SOC is not insensitive to the details of the
system such as the shape of the grains  and  that inertia effects play an
important role in the relaxation dynamics.  The Oslo group introduced  a
``rice-pile'' model to phenomenologically describe their experiments
\cite{Frette2,Sorensen,Christensen}. This model describes the phenomenology of
rice piles presenting SOC behavior. Moreover, further studies showed that it
can be exactly mapped to a model for interface depinning \cite{Paczuski}.
However, the Oslo model does not present a stretched exponential distribution
in any region of its parameter space.

In a recent communication Head and Rodgers \cite{Head} presented a clustered 
based model for granular flow,  which exhibits both stretched exponential and
power law behavior over its parameter range. Through this model they showed
that  power law behavior   appears to be related to the coherent motion of
clusters of grains that slide along the surface, while the stretched
exponential behavior results from a fluid like motion associated with the
rolling of isotropic grains. However, the motion of individual grains
is implicitly incorporated in this model, where the basic
dynamical variables are related to cluster properties. Hence, it is
interesting to investigate in a more microscopic level how the crossover from
power law to stretched exponential behavior  emerges as a consequence of a
change in the transport properties of individual grains.

In this work we present a model where the ability of {\it individual grains} to 
roll a distance $r$ is described by a {\it long-range} rolling probability of
the form

\begin{equation}
\label{P}
P(r)=\left\{ \begin{array}{ll}
	       A/r^\rho & if \;\;\;\;\; 1 \leq r \leq N \\
	       0                & otherwise\\
	      \end{array}
	      \right.
\end{equation}

\noindent where $A$ is a normalization constant, $N$ is the system size, $r=1,2,\ldots,N$ 
and $0\leq \rho\leq\infty$. The parameter $\rho$ controls the average rolling
distance which is expected to depend on the aspect ratio of the grains. The
$\rho\rightarrow\infty$ limit corresponds to a nearest neighbors movement
($P(r)=\delta_{r,1}$), thus describing the case where the grains do not roll.
In the opposite limit $\rho=0$ the grains can move to any site of the pile (or
even drop off in a single jump) with equal probability.  Hence, this limit can
be associated to the idealized situation of perfect spherical grains without
friction. Moreover, studies on different dynamical systems with similar
long-ranged interactions\cite{Gleiser,Tamarit,Cannas} have shown that, except
from some possible rescaling factor, the behavior of almost all the relevant
properties of the systems in the full range $0\leq\rho< 1$ reproduce the
corresponding ones for $\rho=0$. This appears to be related to the behavior of
the first moment of the distribution (\ref{P}), which in the present context
represents the mean rolling distance of the grains. For large values of $N$ we
have that

\begin{equation}
 \left<r\right>\approx \frac{1-\rho}{2-\rho} \frac{N^{2-\rho}-1}{N^{1-\rho}-1} \sim
  \left \{ \begin{array}{cl}
  \frac{(1-\rho)}{(2-\rho)} N &  \mbox{if $\rho < 1$}   \\ \\ \nonumber
  \frac{N}{ln(N)} & \mbox{if $\rho =1$} \\ \\ \nonumber
  \frac{(\rho-1)}{(2-\rho)} N^{2-\rho} & \mbox{if $1 < \rho < 2$} \\ \\ \label{<r>}
  \frac{ln(N)}{N} & \mbox{if $\rho =2$} \\ \\ \nonumber
  \frac{(1-\rho)}{(2-\rho)} & \mbox{if $\rho > 2$} \nonumber
\end{array}
\right. 
\end{equation}
 
\noindent We see that, for $0\leq\rho< 1$, $\left<r\right>$ is of the order
of the  system size, which means that a very large number of grains will drop
off of the system in a single jump, for any system size. Since this behavior
appears to be highly unrealistic we will concentrate our study on the
$\rho\geq 1$ regime .

Our model is based on the Oslo one \cite{Frette2,Sorensen,Christensen} and it is defined as 
follows. 
We consider  a one dimensional lattice of size $N$ ($1 \le i \le N$), each site $i$ having 
associated an integer variable $h(i)$ representing the local height of the
pile. The  
 local slope is then given by  $\sigma(i)=h(i)-h(i+1)$.
The grains enter into the system from the left ($i=1$) and may drop off at the 
rightmost site $i=N+1$, imposing  $h(N+1)=0$ for all times. 
 Every time  the local slope $\sigma(i)$ of a site $i$ 
exceeds a local critical value $\sigma(i) > \sigma_c(i)$ the  topmost grain at site $i$ rolls 
$r$ sites to the right with probability $P(r)$ given by Eq.(\ref{P}).  Then,
the heights of the sites $i$ and $i+r$ are recalculated as $h(i) \rightarrow
h(i)-1$ and $h(i+r)\rightarrow h(i+r)+1$ and the corresponding local slopes
are modified accordingly.
  Each time a grain leaves a column $i$ we assign ita 
 new critical slope $\sigma_c(i)=1$ or $\sigma_c(i)=2$ with equalprobability. This process is
  repeated until all the local slopes satisfy
$\sigma(i) \le \sigma_c(i)$. An avalanche starts when $\sigma(1) >
\sigma_c(1)$ and when it stops (i.e., $\sigma(i) \le \sigma_c(i)$ $\forall i$)
new grains are added until a new avalanche is initiated. Notice that in the
limit $\rho\rightarrow\infty$  the present model reduces to Oslo model.

\begin{center}
\begin{figure}
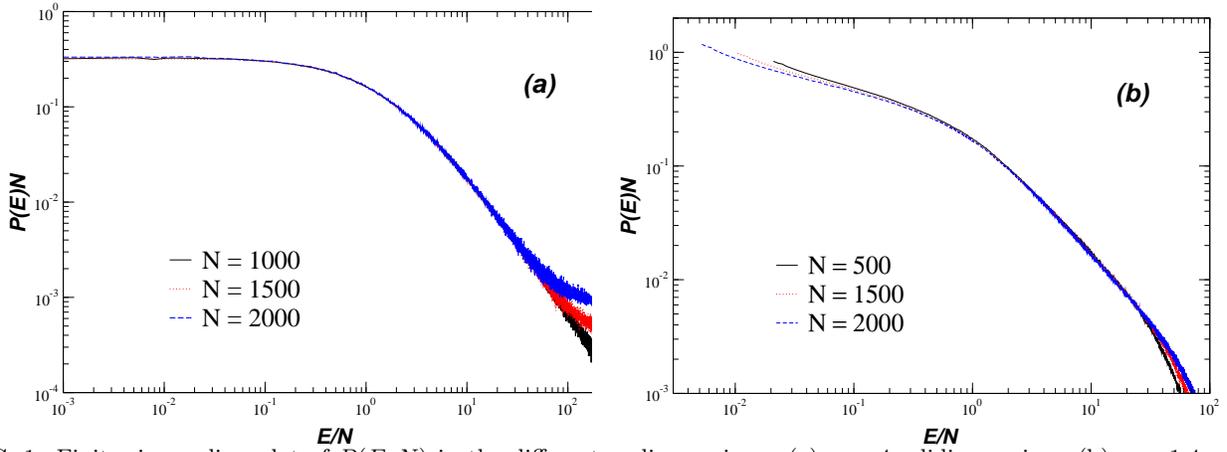

\epsfig{file=fig1a.eps,width=8cm}
\epsfig{file=fig1b.eps,width=8cm}
\caption{Finite-size scaling plot of $P(E,N)$ in the different scaling regimes; (a) $\rho=4$:
 sliding  regime; (b) $\rho=1.4$: rolling regime.}
\label{fig1}
\end{figure}
\end{center}

\begin{center}
\begin{figure}
\epsfig{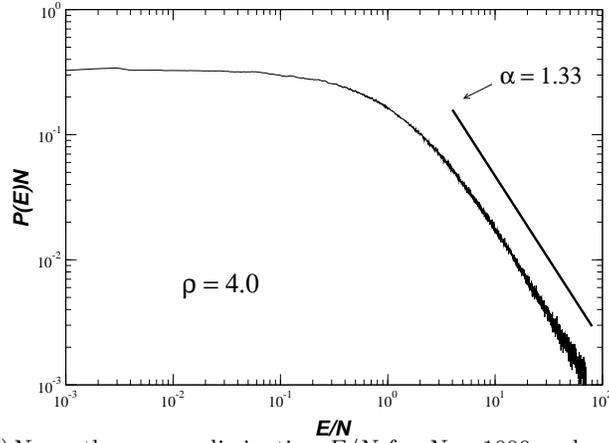}
\caption{Log-log plot of $P(E,N)N$ vs. the energy dissipation $E/N$ for $N=1000$ and $\rho=4$.
 The same qualitative shape is observed in the whole regime $\rho>2$. The
straight line indicates the best fitting in the power law region $f(x)\sim
x^{-\alpha}$.}
\label{fig2}
\end{figure}
\end{center}

Once the system attains the stationary state, fluctuations in the slope of the pile appear 
in the form of avalanches that redistribute the grains, thereby changing the
profile. By considering the profile just before an avalanche is initiated
and the final profile when it stops, the avalanche size  is
defined as the total energy dissipated between both profiles \cite{Frette}.
The local energy change in a site $i$ is calculated as the difference between
the potential energies in both profiles, where the units were chosen such that
$mg=1$. We study for different values of $\rho$ the probability density
$P(E,N)$, $P(E,N)dE$ being the probability that an avalanche with energy
dissipation between $E$ and $E+dE$ will occur in a system of size $N$.
Numerical data were smoothed using a local average procedure in order to 
diminish large statistical fluctuations present for high values of the ratio
$E/N$. We verified  for a wide range of  values of $\rho$ (even in the non
physical region $\rho<1$) that $P(E,N)$ displays the finite-size scaling
behavior $P(E,N)=N^{-1}f(E/N)$, in agreement with the experimental data
\cite{Frette}, as shown by the data collapse of Fig. \ref{fig1}  for
$\rho=1.4$ and $\rho=4$. The scaling function $f(x)$ depends on the value of
$\rho$. We found two clearly distinct regimes. For $\rho>2$ (hereafter
referred as the ``sliding regime'') $f(x)$ is almost constant for small values
of $x$ and presents a power law dependency $f(x)\sim x^{-\alpha}$ for large
values of $x$, as shown in the example of Fig.\ref{fig2} for $\rho=4$. The
exponent $\alpha$ increases smoothly from $\alpha\approx 1$ in the limit
$\rho\rightarrow 2^+$ to $\alpha=1.33 \pm 0.05$ for values of $\rho \geq 4$,
where it becomes independent of $\rho$. This value is consistent with that
obtained by Head and Rodgers \cite{Head} for the cluster based rice pile 
model.

\begin{center}
\begin{figure}
\epsfig{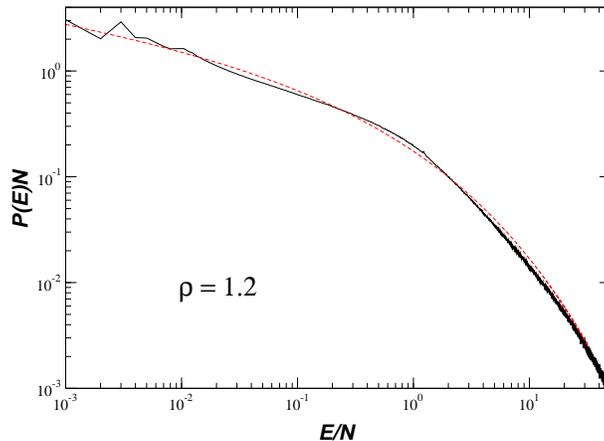}
\caption{Log-log plot of $P(E,N)N$ vs. the energy dissipation $E/N$ for $N=1000$ and $\rho=1.2$.
 The same qualitative behavior is observed in the whole regime $1<\rho<2$. The
dahed line corresponds to the best fitting using a function of the form
$f(x)\propto  x^{-\alpha} exp(-(x/x_0)^\gamma)$ (see table \ref{table1} for
the values of the fitting parameters).}
\label{fig3}
\end{figure}
\end{center}

 For $1<\rho < 2$ (hereafter referred as the ``rolling regime'') the best fitting of
 the numerical data is obtained by a function of the form:

\begin{equation}
f(x)\propto\frac{1}{x^\alpha} \exp{\left(-\left(\frac{x}{x_0}\right)^{\gamma}\right)}
\label{f}
\end{equation}

\noindent as shown in the example of Fig.\ref{fig3} for $\rho=1.2$. The
parameters $\gamma$, $\alpha$ and $x_0$ depend on $\rho$, as shown in table
\ref{table1}, although the small variation of $\gamma$ in this regime suggests
 that it may not depend on $\rho$. Notice that this function seems to describe
the behavior of the {\it whole} distribution, not just for large values of
$E/N$ (where it is dominated by the stretched exponential) but also for small
values of it.

\vspace{0.5cm}

We have shown that the crossover from power law to stretched exponential behavior
observed in the granular dynamics of rice piles can be well described as a
long-range effect resulting from a change in the transport properties of
individual grains. Within this scenario two distinct regimes appear, regarding
the qualitative behavior of the avalanche distribution: a short-range
``sliding'' regime and a long-range ``rolling'' one. A direct inspection of
snapshots of the profiles obtained during the simulations shows that, indeed 
grains group into clusters that move coherently in the sliding regime, while
in the rolling regime they move almost independently. These behaviors are
consistent with what is observed experimentally \cite{Frette} and with Head
and Rodgers results \cite{Head}. Details of these simulations will be
published elsewhere. This change in the transport properties of individual
grains is related to different scaling properties of the average rolling
distance with the system size: while in the rolling regime $d$ scales as a
power law $d \sim N^{2-\rho}$, in the sliding regime $d$ does not depend on
$N$.

In the whole sliding regime $\rho>2$ the system displays the 
qualitative behavior expected for a short-range model, that is, a power law
for large values of the energy density $E/N$ and constant for small values of
it. However, it is worth noting that for intermediate values of $\rho$
($2<\rho < 4$) the characteristic exponent $\alpha$ of the power law
distributions depends on $\rho$, while for large values of $\rho$ it becomes
independent of $\rho$. This results suggests that non-universal behavior may
be expected for intermediate values of the grains aspect ratio. The value
$\rho=4$ is only indicative and it seems not to be a critical value of $\rho$
separating the universal from the non-universal behaviors.

 In the rolling regime the whole avalanche distribution is well described by a
stretched exponential modulated by a power law. This correction to the usual
stretched exponential behavior is only important in the small energy density
region of the spectrum, since for $x \gg x_0$ the function is always dominated
by the stretched exponential. Notice that for $1.4<\rho<2$ the exponent
$\alpha$ changes of sign taking small absolute values, suggesting  logarithmic
 rather than power law corrections to the stretched exponential in this region.
Indeed, a slightly better fitting is obtained with a function of the form

\begin{equation}
f(x) \propto \ln{(x)} \exp{\left(-\left(\frac{x}{x_0}\right)^{\gamma}\right)}
\label{f2}
\end{equation}

\noindent without   any significant change in the fitting parameter values of
$x_0$ and $\gamma$ shown in table \ref{table1}.

From table
\ref{table1} we see that the best agreement \cite{Frette} with the
experimental values  $x_0=0.45 \pm 0.09$ and $\gamma=0.43
\pm 0.03$  is obtained   for  $\rho =1.01$. Both functions Eq.(\ref{f}) and
(\ref{f2}) cease to give good fittings of the simulation data as we approach
the border values $\rho\rightarrow 1^{+}$ and  $\rho\rightarrow 2^{-}$
respectively. This  fact, together   with  an increase in the statistical
fluctuations in the neighborhood of those values, makes very difficult to
 determine how the
crossover from stretched exponential and power law occurs around $\rho=2$.
Possibly the calculation of the distributions of other quantities (like transit
times) may be useful. Some work along this line is in progress.

\newpage

\begin{table}
 \begin{tabular}{|l|r|c|c|} 
\hline 
 $\rho$  & $\alpha$  &  $x_0$  &  $\gamma$  \\ 
\hline 
{$1.01$} & {$0.17 \pm 0.05$} & {$ 0.34 \pm 0.05 $} & {$0.41 \pm 0.05 $} \\  
{$1.1$} & {$0.15 \pm 0.05$} & {$ 0.23 \pm 0.05 $} & {$0.35 \pm 0.05 $} \\ 
{$1.2$} & {$0.15 \pm 0.05$} & {$ 0.20 \pm 0.05 $} & {$0.35 \pm 0.05 $} \\ 
{$1.4$} & {$0.05 \pm 0.05$} & {$ 0.15 \pm 0.05 $} & {$0.33 \pm 0.05 $} \\
{$1.6$} & {$-0.03 \pm 0.05$} & {$ 0.11 \pm 0.05 $} & {$0.31 \pm 0.05 $} \\
{$1.8$} & {$-0.09 \pm 0.05$} & {$ 0.06 \pm 0.05 $} & {$0.28 \pm 0.05 $} \\
 \hline
\end{tabular}
\caption{Fitting parameters using the function (\ref{f}) in the rolling regime $1<\rho< 2$}
\label{table1}
\end{table}

 This work was partially supported by grants from
Consejo Nacional de Investigaciones Cient\'\i ficas y T\'ecnicas CONICET 
(Argentina), Consejo Provincial de Investigaciones Cient\'\i ficas y 
Tecnol\'ogicas (C\'ordoba, Argentina) and  Secretar\'\i a de Ciencia y 
Tecnolog\'\i a de la Universidad Nacional de C\'ordoba (Argentina).

\end{document}